\documentclass[pra,aps,superscriptaddress,notitlepage,longbibliography,twocolumn,nofootinbib,superscriptaddress,10pt]{revtex4-2}
\usepackage{amsmath}
\usepackage{amsfonts}
\usepackage{graphicx}
\usepackage{bm}
\usepackage{bbold}
\usepackage{color}
\usepackage{braket}
\usepackage{comment}
\usepackage{multirow}
\usepackage{amssymb}
\usepackage[english]{babel}
\usepackage[dvipsnames]{xcolor}

\usepackage[colorlinks]{hyperref}
\newcommand{\intinf}{\int_{-\infty}^{\infty}}

\hypersetup{
    colorlinks=true,
    linkcolor=blue,
    filecolor=magenta,      
    urlcolor=magenta,
    citecolor={blue},
    }

\DeclareMathAlphabet{\mathcallc}{U}{dutchcal}{m}{n}
\SetMathAlphabet{\mathcallc}{bold}{U}{dutchcal}{b}{n}
\DeclareMathAlphabet{\mathbcallc}{U}{dutchcal}{b}{n}

\begin{document}

\abovedisplayskip=6pt
\abovedisplayshortskip=6pt
\belowdisplayskip=6pt
\belowdisplayshortskip=6pt

\title{Quantum Optical Soliton Dynamics Beyond Linearization: An Open-System Approach 
}
\author{Chris Gustin}
\email{chris.gustin@nbi.ku.dk}
\altaffiliation[Present address: ]{Center for Hybrid Quantum Networks (Hy-Q), Niels Bohr Institute,
University of Copenhagen, Jagtvej 155A, Copenhagen DK-2200, Denmark}
\affiliation{E.\,L.\,Ginzton Laboratory, Stanford University, Stanford, California 94305, USA}
\author{Ryotatsu Yanagimoto}
\affiliation{Physics \& Informatics Laboratories, NTT Research, Inc., Sunnyvale, California 94085, USA}
\affiliation{School of Applied and Engineering Physics, Cornell University, Ithaca, New York 14853, USA}
\author{Edwin Ng}
\affiliation{Physics \& Informatics Laboratories, NTT Research, Inc., Sunnyvale, California 94085, USA}
\affiliation{E.\,L.\,Ginzton Laboratory, Stanford University, Stanford, California 94305, USA}
\author{Hideo Mabuchi}
\affiliation{E.\,L.\,Ginzton Laboratory, Stanford University, Stanford, California 94305, USA}

\begin{abstract}
We introduce two approaches to modeling the  quantum dynamics of optical $\chi^{(3)}$ solitons. Taking an open-system viewpoint, we project the underlying quantum field into system (soliton) and residual reservoir components. The reservoir is treated as either (i) a discrete ``Lanczos supermode'' (LSM) expansion which  localizes dynamics to a few-supermode basis, or (ii) a  non-local environment which can be traced out by deriving a Markovian master equation (ME). 
Using these methods, we analyze and identify the quantum structure of both the soliton's stability and its hierarchy of perturbations. 
Through numerical simulations, we confirm both methods effectively capture quantum-induced soliton phase shifts in a concise few-mode (single-mode for ME) basis, and the LSM approach also captures photon loss which arises from non-Markovian dispersive couplings. As neither method is limited to the linearized regime, our approaches provide powerful computational tools to analyze complex non-Gaussian quantum dynamics of solitons where other commonly-used  methods fail, providing insight into such non-perturbative regimes. We also investigate radiation that occurs in the presence of higher-order dispersion with ultrashort pulses, deriving a ME that predicts photon loss consistent with classical theory, but find that both classical and ME theory dramatically underestimate the actual amount of dissipation, which we explain in terms of dispersive coupling-induced soliton broadening.

\end{abstract}

\maketitle

\section{Introduction}
Solitons are a widespread and remarkable consequence of nonlinear dynamics wherein certain waveforms retain their shape as they propagate through a balancing of linear and nonlinear effects. Solitons can manifest in a wide range of physical systems, including shallow water~\cite{Hirota1976}, Bose-Einstein condensates (BECs)~\cite{Khaykovich2002}, plasma~\cite{Lonngren1983}, gravity~\cite{verdaguer}, and nonlinear optics~\cite{Kivshar2003}. Unique properties of solitons are utilized in applications (and proposed applications) including interferometry~\cite{McDonald2014,Tsarev2018}, communication~\cite{Marin-Palomo2017}, microscopy~\cite{Andresen2011}, and frequency metrology~\cite{Shen2020}. Beyond classical physics,  quantum mechanical aspects of solitons as ``particles'' have been extensively studied in the context of quantum field theory~\cite{Weinberg2012,Jackiw1977,Maki1986}.

Nonlinear optics is a platform in which quantum dynamics of solitons have been widely studied both from theoretical and experimental standpoints~\cite{Drummond1993}. Notable non-classical features of optical solitons include Gordon-Haus timing jitter~\cite{Gordon1986}, soliton squeezing~\cite{Carter1987,Rosenbluh1990}, quantum soliton evaporation~\cite{Villari2018}, and non-classical phase-space dynamics~\cite{Korolkova2001,Yanagimoto2020,Yanagimoto2021Oct}, to name a few. 
The question of whether these quantum effects are necessarily detrimental or could be harnessed for potential applications could have
significant implications for quantum information science and engineering~\cite{Tsang2006,Yanagimoto2020}. Recent  observations of quantum-induced soliton diffusion~\cite{Bao2021, Jeong2020, Jia2020, Erkintalo2021} highlight significant progress in experimental technologies on this front, underscoring the need for furthering our understanding of quantum soliton dynamics.

In particular, optical solitons are typically identified with classically stable waveforms which form exact solutions to Maxwell's equations in a nonlinear medium with dispersion and bulk nonlinearity. However, \emph{intrinsic quantum fluctuations} will generally perturb such a solution away from its stable point, leading to modifications of soliton phase~\cite{Wright1991,PhysRevA.52.4871,PhysRevA.48.2361,Korolkova2001}, and even dissipation of the waveform itself~\cite{PhysRevA.107.053513,PhysRevA.98.043859,sangshin2025} (soliton ``evaporation'') on a timescale determined by the strength of the nonlinearity.

One very common set of approaches to quantum soliton dynamics under the nonlinear Schr\"odinger equation (NLSE) are linearized treatments~\cite{Villari2018,Helt2020,Haus1990,Matsko2013,PhysRevA.107.053513,sangshin2025}, where quantum fluctuations around the mean fields are approximated perturbatively as Gaussian~\cite{Olivares2012,Braunstein2005}. Unraveling the physics of solitons in non-perturbative regimes where quantum fluctuations are intrinsically \emph{non-Gaussian}, however, can be more complicated, and requires an approach that goes beyond linearization. 
One commonly used approach is the time-dependent Hartree (TDH) approximation~\cite{Yoon1977,Calogero1975,Haus1989, Wright1991,Levit1980,Singer1992}. In TDH, an ansatz with a common wavefunction for every photon is solved self-consistently, which is closely related to the exact many-body eigenstates~\cite{Lieb1963} in the large excitation limit~\cite{Yoon1977,Calogero1975}. However, the uncontrolled nature of the TDH approximation  makes it challenging to account for higher-order contributions, and provides limited insight into the source of the quantum perturbations of the classically stable waveform. More troublesomely, we show in this work~\cite{SI} that the TDH approximation as usually applied rests on an unjustified mathematical approximation, properly accounting for which gives results which do not recover the classical limit as the number of initial soliton photons becomes large. Another approach for $\chi^{(3)}$ optical solitons is to use the exact eigenstates of the field Hamiltonian~\cite{PhysRevA.40.854,PhysRevA.52.4871,PhysRevA.52.1574,PhysRevA.53.454,Villari2018}; however, such an approach is not easily generalizable 
due to the difficulty of finding exact many-body eigenstates for arbitrary Hamiltonians.

The potential complexity of quantum soliton dynamics renders them an ideal platform for studying many of the more challenging and fascinating aspects of quantum nonlinear optics which have received interest in recent years~\cite{Yanagimoto2024Jul,Jankowski:24}; the balancing of nonlinear and linear interactions renders solitons an intrinsically \emph{non-perturbative} phenomenon with respect to the different interactions involved, and the dynamics are generally multimode and can contain many photons. In particular, in the so-called ``mesoscopic''~\cite{Yanagimoto2024Jul,Jankowski:24} regime where the number of photons constituting the soliton is substantial (i.e., dozens-to-hundreds) but still few enough to exhibit significant non-Gaussian dynamics, traditional models are strongly challenged by the very large Hilbert spaces upon which the quantum soliton propagates. Nonetheless, the presence of a stable classical (mean-field) solution raises exciting prospects for using ideas inspired by open quantum system theory~\cite{Gustin2025Feb} as a means to extract crucial non-perturbative information about the underlying quantum state while simultaneously strongly truncating the size of the Hilbert space needed to describe the full dynamics. Namely, we take the viewpoint in this work that the soliton can be viewed as a quantum ``system,'' coupled to a residual environment (or ``reservoir'') which exerts only a weak influence upon the quantum soliton system, through couplings which can be deduced entirely \emph{ab initio}. In other words, a soliton can be seen as a flying temporal ``cavity'' within this picture.

\begin{figure}[thb]
\includegraphics[width=1\columnwidth]{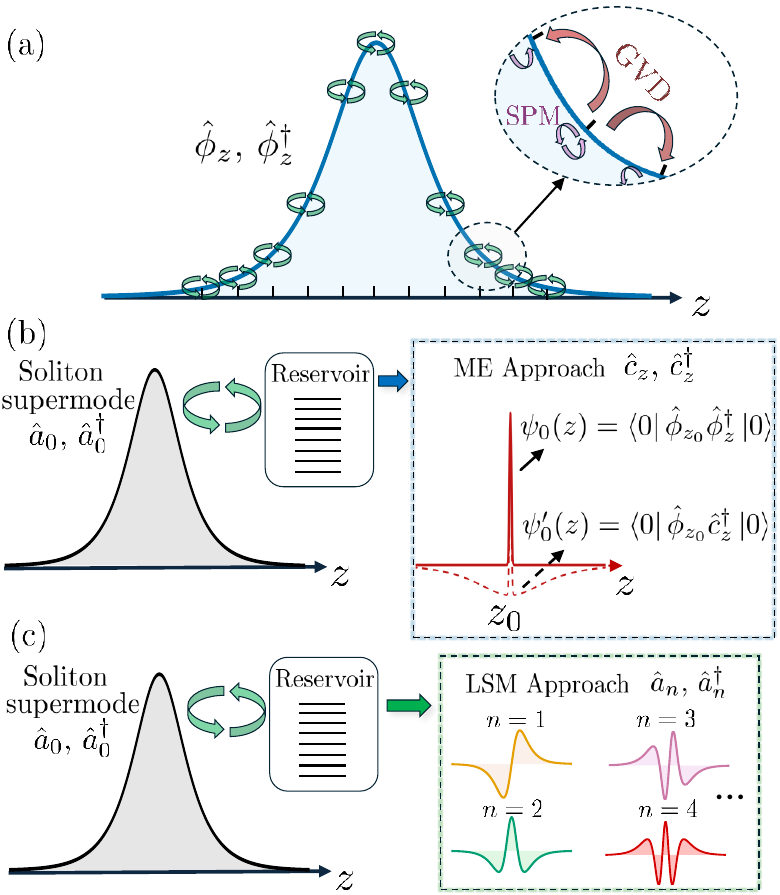}
   \caption{Schematic of quantum soliton (a) continuum model described by Eq.~\eqref{eq:H_fund_1} and (b,c) our reduced approaches using the ME and LSM formalisms, respectively. In (a), the zoom-in shows the self-phase modulation (SPM) self-interactions and group velocity dispersion (GVD) local interactions experienced at each position $z$ along the waveform. In (b,c), the soliton subspace couples to an effective reservoir, which can be represented either as a (b) non-local continuum (ME approach) or (c) as a finite set of LSMs. The ME continuum operators $\hat{c}_{z}^{\dagger}$ create spatially distributed single-photon wavefunctions (with a soliton ``hole'') $\psi'_0(z) = \delta(z-z_0) - u_0(z_0)u_0(z)$. In (c), we show the first four LSMs $u_n(z)$ with $n \geq 1$ (for odd $n$ they are imaginary so we plot the imaginary part).
    }\label{fig:schem}
\end{figure}

Specifically, in this work we introduce two novel methods to study quantum $\chi^{(3)}$ soliton propagation: Lanczos supermodes (LSMs), and a quantum master equation (ME) with a non-local reservoir. In both approaches, we take the quantum soliton to be our system, and then expand the dynamics of the residual quantum reservoir using either a LSM or continuum (which is then traced out using ME techniques) approach. Figure~\ref{fig:schem} shows a schematic of this process. In addition to offering physical insight into quantum soliton stability and perturbations, both of our methods are capable of efficiently modeling challenging non-Gaussian regimes where existing linearization-based techniques fail.
While we base our discussions on optical $\chi^{(3)}$ solitons for concreteness, our formalism is general and can be applied to a range of bosonic systems, including BECs, and $\chi^{(2)}$ nonlinear waveguides with quadratic solitons.

LSMs are a supermode basis~\cite{Brecht2015, Onodera2022Mar, Patera2010} specifically designed to concisely model quantum mechanical behaviors of solitons. In our approach, we orthonormalize a Krylov function subspace composed of waveforms obtained via iterative applications of the momentum operator to the classical soliton waveform. By employing a Lanczos iteration for the orthonormalization procedure~\cite{Lanczos1950}, we naturally ensure the locality of the interactions among supermodes induced by dispersion, thus constraining dynamics in a few-supermode basis, and concisely capturing various quantum features of an optical soliton. For instance, the soliton stability can be seen as a suppression of leakage from the fundamental supermode due to destructive interference between linear and nonlinear couplings. 

We also introduce a quantum ME method for analyzing quantum soliton propagation. The method is based on partitioning the underlying continuum field into a fundamental soliton supermode, and a residual (continuum) reservoir component, which by nature of its construction is intrinsically \emph{non-local}, in the sense that its fundamental excitations are labeled by a continuous index, but distributed over space. Using Markovian ME techniques, we trace out the non-local reservoir degrees of freedom, and show that this approach can concisely capture quantum perturbations of the classical soliton phase evolution while retaining the simplicity of a system corresponding to a single field supermode. Interestingly, we find that coupling channels that initially seem non-perturbative---and thus, non-Markovian---become Markovian due to effective destructive interference between dispersive and nonlinear interaction pathways. The result is a single-mode effective ME (or indeed, as we show, often an effective \emph{Hamiltonian}) model of the quantum soliton, which resides in a vastly smaller Hilbert space than the initial quantum field description.

In addition to its application herein to quantum soliton dynamics, our ME formalism with a projected non-local reservoir is highly general and has significant prospects for application in a broad range of quantum optical platforms. It has recently been used, for example, in cavity-QED to derive corrections to MEs in broadband coupling regimes~\cite{gustin2025dissipationbroadbandultrastrongcoupling} and to model cavities coupled with time-delayed photons~\cite{fuchs2026quantumdynamicscoupledquasinormal} through the quantization of intrinsically lossy cavity modes~\cite{Franke2019May}, and bears connections to reaction coordinate~\cite{garg_effect_1985,shubrook_non-markovian_2025,PhysRevA.104.052617,noauthor_effective-mode_nodate} and pseudomode~\cite{PhysRevA.50.3650,PhysRevA.55.2290,PhysRevA.64.053813,PhysRevResearch.2.043058,Lambert2019Aug} theories. This work extends its application to a system with \emph{nonlinear} system-reservoir interactions (as well as a nonlinear reservoir Hamiltonian), and we show that it remains robust as a method of deducing open quantum system dynamics from an \emph{ab initio} approach.

The layout of the rest of the paper is as follows: In Sec.~\ref{sec:theory} we briefly summarize elements of classical $\chi^{(3)}$ theory and the quantum field Hamiltonian which we base our work upon, then present our LSM and ME formalisms. In Sec.~\ref{sec:results_soliton}, we describe the hierarchy of terms which contribute to the soliton stability and quantum fluctuations, then present numerical simulations in the low and high photon number regimes to validate our LSM and ME theories by comparison with numerically-exact matrix product state (MPS) and Gaussian split-step Fourier (GSSF) methods~\cite{ng2023quantumnoisedynamicsnonlinear}. In Sec.~\ref{sec:higher-order}, we investigate the effect of considering third-order dispersion (TOD), which manifests in the ME theory as a set of nonlinear dissipation Lindblad operators and corresponding dispersive Hamiltonian corrections. We find that while these recover classical predictions, they underestimate the actual amount of dissipation observed, and we provide an explanation for this observed effect in terms of dispersive spreading of the soliton profile enhancing the dissipation rate. Finally, in Sec.~\ref{sec:conc_outlook}, we conclude and give outlook for potential experimental connections and further work.

We also include a Supplementary Information (SI)~\cite{SI}, which contains the nondimensionalization of our starting point field Hamiltonian, further mathematical details of our LSM and ME formalisms and simulation methods, and a derivation of how the TDH theory, as often applied to soliton analysis, fails to predict quantum fluctuations consistent with the known classical limit.

 \section{Theoretical Methods}\label{sec:theory}
 In this section, we give an overview of the fundamental quantum Hamiltonian model we use to describe soliton propagation, as well as our LSM and ME formalisms, which are then used in later sections to study various phenomena. Detailed mathematical details can be found in~\cite{SI}.
\subsection{Quantum Model and Classical solitons}\label{sec:field}
Our model of the quantum  $\chi^{(3)}$ dynamics in a nonlinear waveguide begins~\cite{SI} with the Hamiltonian
\begin{equation}\label{eq:H_fund_1}
    \hat{H} = -\frac{1}{2}\int_{-\infty}^{\infty} dz \hat{\phi}^{\dagger}_z \partial^2_z \hat{\phi}_z - \frac{1}{2}\int_{-\infty}^{\infty} dz \hat{\phi}^{\dagger 2}_z \hat{\phi}^2_z +\hat{H}_N
\end{equation}
where $\hat{\phi}_z$, $\hat{\phi}^{\dagger}_z$ describe annihilation and creation operators of photons in a co-moving frame at position $z$ along the waveguide, which satisfy $[\hat{\phi}_z,\hat{\phi}^{\dagger}_{z'}] = \delta(z-z')$. The first term in Eq.~\eqref{eq:H_fund_1} corresponds to a quadratic dispersion relationship, and the second term gives a third-order (Kerr) nonlinear optical nonlinearity. The term $
    \hat{H}_N = \alpha_N\intinf dz \hat{\phi}_z^{\dagger} \partial^N_z \hat{\phi}_z$ 
represents higher ($N^{\rm th}$) order dispersion; henceforth we let $\alpha_N=0$ (except for in Sec.~\ref{sec:higher-order} where we study dispersive wave radiation), while the SI~\cite{SI} treats the general case.

The Hamiltonian $\hat{H}$ generates the equation of motion for the field coordinate $
    i\partial_t \hat{\phi}_z = -\frac{1}{2}\partial^2_z \hat{\phi}_z - \hat{\phi}^{\dagger}_z \hat{\phi}^2_z$,
which, in the mean-field limit where $\hat{\phi}_z \rightarrow \phi_z$, is the classical nonlinear Schr{\"o}dinger equation (NLSE). 
The NLSE has a family of soliton solutions $\phi_z(t) = f_{\bar{n}}(z,t)$, indexed by $\bar{n}$, given by
\begin{equation}\label{eq:fund_soliton}
     f_{\bar{n}}(z,t) =\frac{\bar{n}}{2}\text{sech}\left(\frac{\bar{n}z}{2}\right) e^{i\frac{\bar{n}^2}{8}t},
\end{equation}
where $\bar{n}$ is the number of photons in this classical fundamental soliton solution. It is convenient to define $T_0 = 2\pi/\bar{n}^2$, the fundamental soliton period~\cite{Kivshar2003}.

Our interest is in understanding the quantum dynamics of the soliton---that is, how does the state inside the soliton pulse shape evolve? As the classical soliton waveform does not change as a function of time, it seems natural to assume the \emph{quantum} dynamics are also near single-moded. This motivates us to project the quantum dynamics onto a mode with spatial profile following the classical soliton. Thus, we can consider writing the quantum Heisenberg picture field operator as $\hat{\phi}_z(t) = u_0(z)\hat{a}_0(t)$, where $
    \hat{a}_0 = \int_{-\infty}^{\infty} dz u_0(z) \hat{\phi}_z$,
and $u_0(z) = \bar{n}^{-\frac{1}{2}}f_{\bar{n}}(z,t=0)$, such that $\hat{a}_0$ is a ``soliton supermode'' Fock space operator that satisfies $[\hat{a}_0,\hat{a}_0^{\dagger}]=1$. Expressing Eq.~\eqref{eq:H_fund_1} in this single-mode basis, we thus find
\begin{equation}
    \hat{H}_{0} = \frac{\bar{n}^2}{24}\hat{a}_0^{\dagger}\hat{a}_0 - \frac{\bar{n}}{12}\hat{a}_0^{\dagger 2}\hat{a}_0^2.
\end{equation}
However, this decomposition is not a complete basis of expansion for the field coordinate; while $\hat{H}_0$ generates the classical solution $\langle \hat{a}_0 \rangle(t) = \langle \hat{a}_0 \rangle (0) e^{i\bar{n}^2 t/8}$ in the mean-field limit, it does not fully accurately predict quantum corrections to this solution, as we confirm in Sec.~\ref{sec:results_soliton}. The central perspective of our work is thus to expand the full field operator as $\hat{\phi}_z = \mathcal{P}\hat{\phi}_z + \mathcal{Q}\hat{\phi}_z$, where $\mathcal{P}$ and $\mathcal{Q}$ are projection superoperators which satisfy $\mathcal{P}+\mathcal{Q}=1$, where $\mathcal{P}\hat{\phi}_z = u_0(z)\hat{a}_0$ defines the soliton ``system'' component, and the residual component $\mathcal{Q}\hat{\phi}_z$ can be treated as a  ``reservoir'' for the fundamental soliton. We introduce two complementary approaches to treat $\mathcal{Q}\hat{\phi}_z$---one based on a discrete expansion in terms of judiciously chosen LSMs, and one based on a continuum non-local reservoir from which we derive a quantum ME. The following subsections give a brief outline of these methods; mathematical details can be found in the SI~\cite{SI}.

\subsection{LSM approach}
\begin{figure}[bth]
\includegraphics[width=0.93\columnwidth]{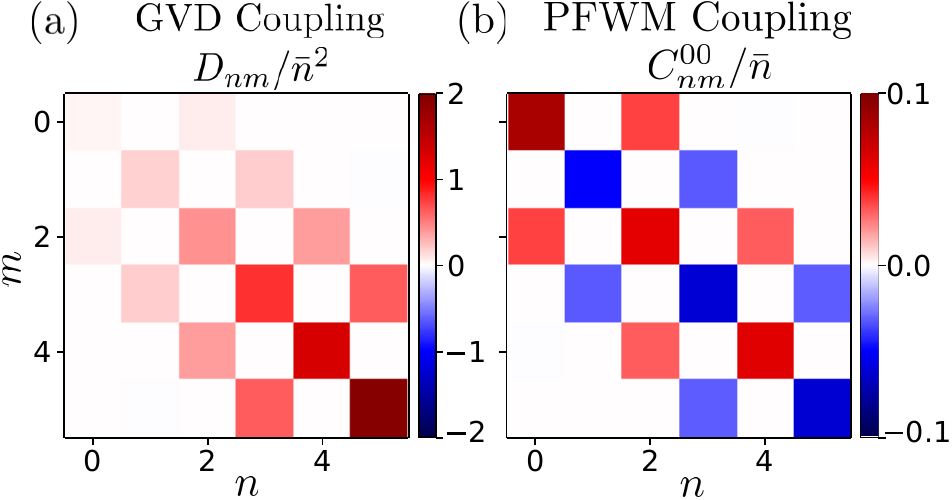}
    \caption{LSM coupling matrices for (a) GVD-induced linear coupling $D_{nm}$ and (b) PFWM $C_{nm}^{00}$.}
     \label{fig:coupling}
 \end{figure}

In the LSM approach, we define the ``reservoir'' superoperator projection to satisfy
\begin{equation}\label{eq:Q_LSM}
    \mathcal{Q}_{\rm LSM}\hat{\phi}_z = \sum_{n=1}^{N_{\rm LSM}-1} u_n(z) \hat{a}_n,
\end{equation}
where $u_n(z)$ are the $N_{\rm LSM}$ LSM profiles, defined such that $[\hat{a}_n,\hat{a}_m^{\dagger}]=\delta_{nm}$. The construction of these LSMs begins with the Krylov subspace $\{u_0(z),Au_0(z),...,A^{N_{\rm LSM}-1}u_0(z)\}$, which is generated by a Hermitian (function space) operator $A$. As the elements of the Krylov subspace are generically linearly dependent, we perform an orthonormalization scheme by means of the Lanczos iteration method~\cite{Lanczos1950,Park1986}, which can be used to ensure locality of couplings when the Hamiltonian~\eqref{eq:H_fund_1} is expressed in the LSM basis and the Hermitian operator $A$ is chosen to represent a dispersive interaction by means of the momentum operator $A = -i\partial_z$~\cite{SI}. This procedure generates an orthonormal subspace of LSMs described by $u_n(z)$---some are shown in Fig.~\ref{fig:schem}(c)---which can be used in the decomposition of Eq.~\eqref{eq:Q_LSM} to express the full Hamiltonian~\eqref{eq:H_fund_1} in the LSM basis:
\begin{equation}
    \hat{H} = \sum_{nm}D_{nm}\hat{a}_n^{\dagger}\hat{a}_m - \sum_{nmkl}C_{nm}^{kl}\hat{a}_n^{\dagger}\hat{a}_m^{\dagger}\hat{a}_{k}\hat{a}_{l},
\end{equation}
where $D_{nm} = -\frac{1}{2}\intinf dz u_n^*(z)\partial^2_zu_m(z)$ and $C_{nm}^{kl} =\frac{1}{2} \intinf dz u_n^*(z)u_m^*(z)u_k(z)u_l(z)$ are tensors representing the linear group velocity dispersion (GVD) and nonlinear $\chi^{(3)}$ parametric four-wave mixing (PFWM) interactions. For all elements, $D_{nm} \sim \bar{n}^2$ and $C_{nm}^{kl} \sim \bar{n}$, respectively. Our choice of LSMs ensures the linear term of the Hamiltonian (and part of the nonlinear term) has a \emph{local} coupling structure, as shown in Fig.~\ref{fig:coupling}.

Crucially, the structure embedded in the $\chi^{(3)}$ Hamiltonian which enables the existence of soliton states is reflected in our LSM formalism as a cancellation between linear and nonlinear coupling terms between the soliton supermode, $n=0$, and the first subdominant supermode of the same parity, $n=2$, which manifests as stability of the quantum soliton, as we show in Sec.~\ref{sec:instability}. This cancellation, in addition to the local structure of the linear couplings, ensures that the LSM basis allows for concise modeling of soliton dynamics, as the key features of the full quantum evolution can be captured by use of just a few LSMs.

\subsection{ME Approach}
In the ME approach, we instead define the reservoir superoperator projection as $\mathcal{Q}_{\rm ME} \hat{\phi}_z = \hat{c}_z$, where $\hat{c}_z$ is an annihilation operator corresponding to the orthogonal complement of the supermode subspace, defined in terms of a \emph{continuous} coordinate $z$. This allows us to ``pull out'' the discrete fundamental supermode from the continuum, such that the system and reservoir subspaces together span the same total space as that of the original fundamental field operators $\hat{\phi}_z$, $\hat{\phi}_z^{\dagger}$. The operators of this complement subspace commute with those of the soliton $\hat{a}_0$, $\hat{a}_0^{\dagger}$, and themselves have non-local commutation relation
\begin{equation}\label{eq:comm}
[\hat{c}_{z},\hat{c}_{z'}^{\dagger}] = \delta(z-z') - u_0(z)u_0(z').
\end{equation}
These commutation relations also define the photon wavefunctions of the fundamental excitations of the reservoir field, as shown in Fig.~\ref{fig:schem}(b). We can interpret such nonlocality as arising from the requirement that the reservoir field is quantized subject to the constraint that its excitations remain orthogonal to the soliton subspace, similar to how the commutation relations of the canonical electromagnetic field form a transverse delta function to enforce the constraint of transverse quantized fields~\cite{Cohen1997}.
By expressing the Hamiltonian $\hat{H}$ in terms of supermode and reservoir operators, we can construct a system-reservoir Hamiltonian $\hat{H} = \hat{H}_{\rm S} + \hat{H}_{\rm R} + \hat{H}_{\rm S-R}$, with system, reservoir, and interaction terms given by
\begin{subequations}\label{eq:H_SR}
\begin{align}
    \hat{H}_{\rm S} &= \hat{H}_0 \\
    \hat{H}_{\rm R} &= -\frac{1}{2}\int_{-\infty}^{\infty} dz \hat{c}^{\dagger}_z \partial^2_z \hat{c}_z - \frac{1}{2}\int_{-\infty}^{\infty} dz \hat{c}^{\dagger 2}_z \hat{c}^2_z \\
    \hat{H}_{\rm S-R} &= \hat{V}_1 + \hat{V}_2 + \hat{V}_{\rm others},
\end{align}
\end{subequations}
where $\hat{V}_1 = \hat{B}_{\rm L}\hat{a}_0^{\dagger}(1-\hat{a}_0^{\dagger}\hat{a}_0/\bar{n}) + \text{H.c.}$, and $\hat{V}_2 = \hat{B}_{\rm NL}\hat{a}_0^{\dagger 2} + \text{H.c.}$, with $\hat{B}_{\rm L} = -\frac{1}{2}\intinf dz \partial^2_z u_0(z) \hat{c}_z$, and $\hat{B}_{\rm NL} =- \frac{1}{2}\intinf dz u_0^2(z) \hat{c}_z^2$. $\hat{V}_{\rm others}$ consists of terms which annihilate the vacuum and subsequently do not contribute under the Markov approximation we will make.

To derive a ME, we employ the standard 2nd order Born-Markov formalism with the usual assumptions~\cite{breuer2002theory}---most importantly, that the reservoir begins in vacuum and stays close to it over time. An important aspect of our formalism is that due to the non-local commutation relation of Eq.~\eqref{eq:comm}, the interaction picture transformation of the reservoir operators (necessary to derive the ME) takes a non-trivial form in the time domain. By means of the convolution theorem, the non-local reservoir Green functions that enter into the final dissipator in the ME can nonetheless be evaluated in closed form~\cite{SI}. This simplification applies only to the Green function generated by the 1-reservoir-photon coupling $\hat{V}_1$---however, in the SI, we make use of the exact diagonalizability of $\hat{H}_{\rm R}$ to evaluate those arising from the 2-reservoir-photon coupling $\hat{V}_2$ contribution as well~\cite{SI}.

Upon evaluation of the Born-Markov ME, we find that the ME for the soliton subspace (with reduced density operator $\hat{\rho}$) takes the form $\partial_t\hat{\rho} = -i[\hat{H}_0,\hat{\rho}] + \mathcal{L}_1\hat{\rho} + \mathcal{L}_2 \hat{\rho}$, where the 1-reservoir-photon and 2-reservoir-photon dissipators take the form
\begin{subequations}\label{eq:dissipators}
\begin{equation}\label{eq:F}
\mathcal{L}_{1}\hat{\rho} = \sum_{n,m=1}\Lambda_{nm}[\hat{\sigma}_{n} \hat{\rho}, \hat{\sigma}_{m}^{\dagger}]
 + \text{H.c.},
 \end{equation}
 \begin{equation}
     \mathcal{L}_2 \hat{\rho} = \sum_{n,m=2} \Gamma_{nm}[\hat{s}_n,\hat{\rho},\hat{s}_m^{\dagger}] + \text{H.c.},
 \end{equation}
 \end{subequations}
 which are expressed in terms of Fock state transition operators $\hat{\sigma}_n = \ket{n-1}\bra{n}$ and $\hat{s}_n = \ket{n-2}\bra{n}$. The matrix $\Lambda_{nm}$ characterizes the soliton dissipator associated with $\mathcal{L}_1$ in the Fock 1-photon transition basis, and is
\begin{equation}\label{eq:nodisp}
\Lambda_{nm} = -i\frac{\sqrt{nm}}{T_0}\left[1-\frac{n-1}{\bar{n}}\right]\left[1-\frac{m-1}{\bar{n}}\right]R(\tilde{\omega}_n),
\end{equation}
where $\tilde{\omega}_n = \pi T_0(E_n - E_{n-1})$, where $E_n = \frac{\bar{n}^2n}{24} - \frac{\bar{n}(n^2-n)}{12}$ is the eigenenergy of $\hat{H}_0$ for the $n$-photon Fock state, and  
$R(\omega)$ is an analytic function derived in the SI~\cite{SI}. The $\mathcal{L}_1$ contribution to the ME is \emph{stationary} for Fock states with density matrix $\ket{\bar{n}+1}\bra{\bar{n}+1}$, which is a manifestation of the stability of the quantum soliton, as discussed in Sec.~\ref{sec:instability}. It is important to note that $\Lambda_{nm}$ is typically entirely imaginary around $n,m \approx \bar{n}$, and thus no significant dissipation (soliton photon loss) is predicted from the ME theory; see discussion in following section. The 2-reservoir-photon dissipator $\mathcal{L}_2$ is expressed in terms of the rate matrix $\Gamma_{nm}$ which is given in the SI~\cite{SI}. Similarly, the elements of $\Gamma_{nm}$ are almost entirely imaginary and produce little to no photon loss. The ME is in non-Lindblad form but could be recast in Lindblad form if desired (e.g., for stochastic Schr{\"o}dinger simulations)~\cite{Nathan2020Sep}.
    
\begin{figure}[tb]
\includegraphics[width=1\columnwidth]{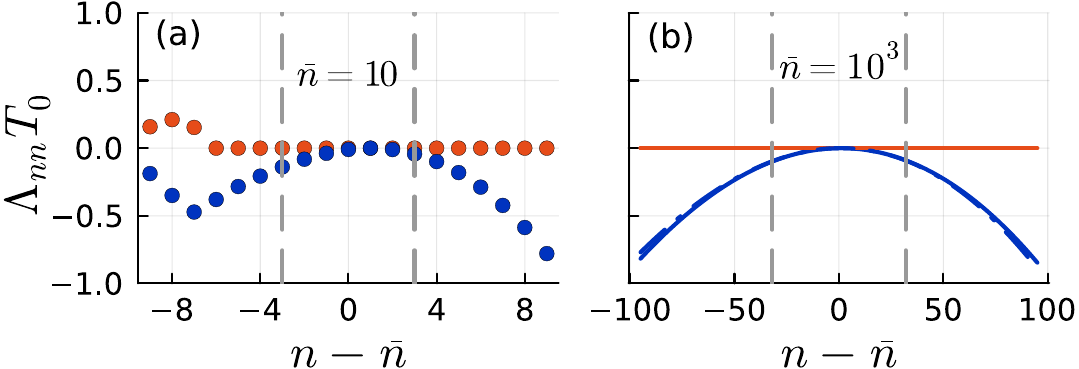}
    \caption{Real (red) and imaginary (blue) components of the diagonal of the ME dissipator matrix $\Lambda_{nn}$ for (a) $\bar{n}=10$ and (b) $\bar{n} = 1000$. In (b) the approximate form which gives the Hamiltonian $\hat{H}_{\rm eff} = \hat{H}_0 +\hat{H}_{\rm fluc}$ is shown as dashed lines (omitting the $\Omega_{\bar{n}}$ contribution, which arises from $\mathcal{L}_2$). The dashed gray lines specify the bounds of one standard deviation $\pm \sqrt{{\bar{n}}}$ of the initial coherent state photon distribution.
    }\label{fig:me_rates}
\end{figure}

    When $\bar{n} \gg 1$, the dissipator term can be written in Hamiltonian form~\cite{SI} $(\mathcal{L}_1+\mathcal{L}_2)\hat{\rho} = -i[\hat{H}_{\rm fluc},\hat{\rho}]$, where
    \begin{equation}\label{eq:Hfluc}
       \hat{H}_{\rm fluc}= \frac{\bar{n}^2}{2\pi^2}\left[3 - \frac{\pi^2}{3}\right]\hat{a}_0^{\dagger}\left[1-\frac{\hat{a}_0^{\dagger } \hat{a}_0}{\bar{n}} \right]^2\hat{a}_0  + \Omega_{\bar{n}}\hat{a}_0^{\dagger 2}\hat{a}_0^2,
    \end{equation}
     where $\Omega_{\bar{n}} \approx -0.017$, such that the soliton evolves with \emph{effective single-supermode} Hamiltonian $\hat{H}_{\rm eff} = \hat{H}_0 + \hat{H}_{\rm fluc}$. In Fig.~\ref{fig:me_rates}, we show the diagonal elements $\Lambda_{nn}$, as well as those corresponding to the approximate Hamiltonian form for the $\mathcal{L}_1$ contribution.

\section{ Results and Analysis}\label{sec:results_soliton}
\subsection{Sources of Quantum Instabilities}\label{sec:instability}
Using the LSM and ME formalisms, we can gain analytical insight into the hierarchy of sources of destabilization for the soliton that arise due to quantum fluctuations. We then support these insights numerically in Sec.~\ref{sec:numerics}. In general, we will assume an initial coherent state,
\begin{equation}\label{eq:CS}
\ket{\psi_{\rm CS}} = e^{\sqrt{\bar{n}}(\hat{a}_0^{\dagger}-\hat{a}_0)}\ket{0},
\end{equation}
which contains $\bar{n}$ photons and is a close analogue of the classical soliton solution. We remark that there are also other definitions of initial ``quantum soliton'' states that can be posited, based on superpositions of exact eigenstates of the $\chi^{(3)}$ Hamiltonian~\cite{PhysRevA.52.4871,PhysRevA.48.2361,Lai1989Jul2,Villari2018}. Both forms recover the classical limit as $\bar{n}\rightarrow \infty$. This section uses general scaling arguments, valid for short time evolution such that the system remains close to a soliton state.

\begin{figure}[ht]
    \centering
    \includegraphics[width=0.99\columnwidth]{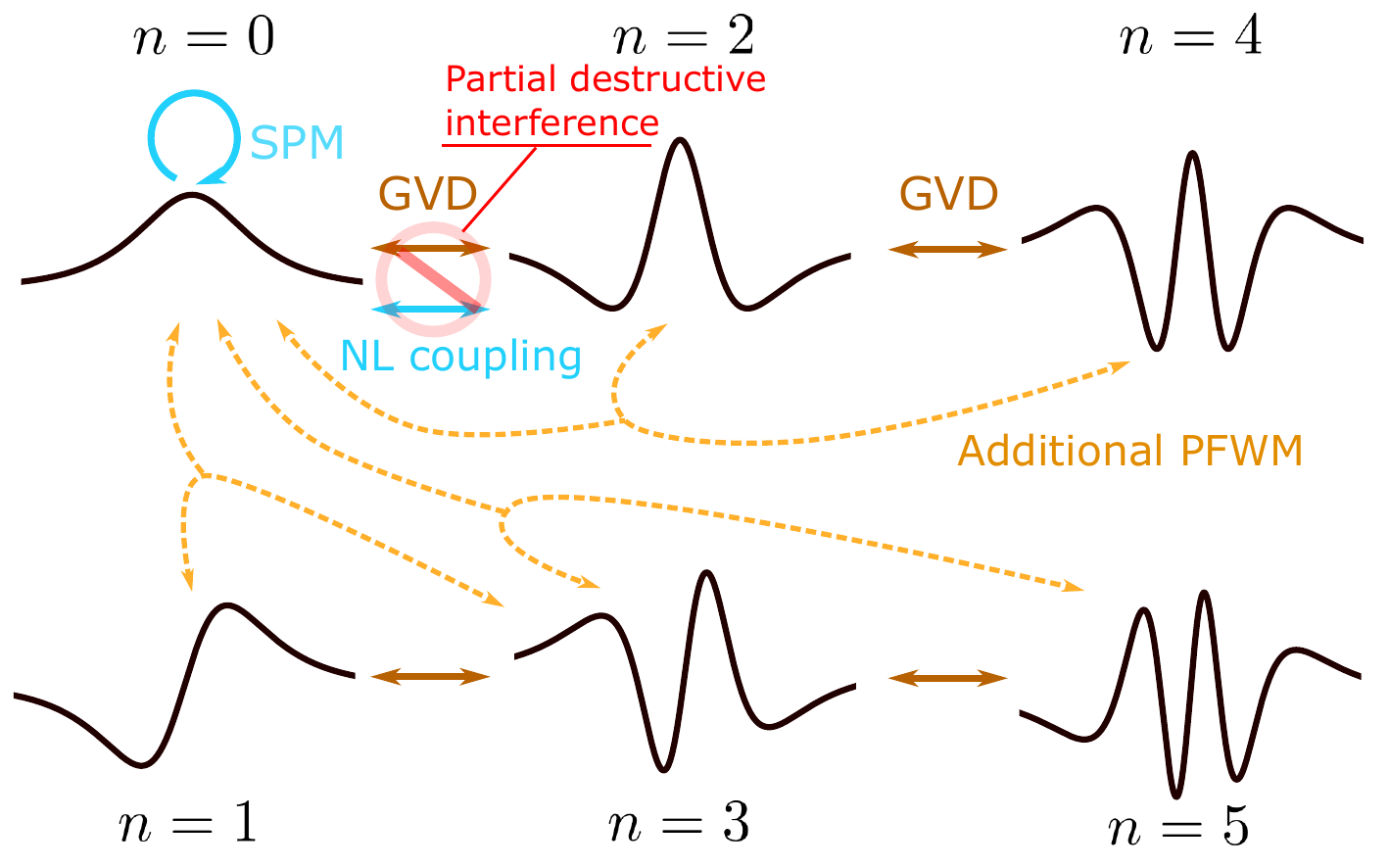}
    \caption{Illustration of LSM interactions. Dominant GVD-induced linear
couplings locally connect supermodes with same parity. In the mean-field, the nonlinear (in the soliton supermode operators) coupling between $n = 0$ (soliton)
and $n = 2$ LSMs  destructively interferes with the GVD-induced coupling to give $\hat{V}_1$. Quantum-induced soliton perturbation and evaporation via PFWM is dominantly through the 2-reservoir-photon coupling $\hat{V}_2$ as well as through higher-order $\hat{V}_{\rm others}$, terms which are negligible for larger $\bar{n}$.
    }\label{fig:supermodes}
\end{figure}

The fundamental stability of the soliton in the quantum mechanical picture arises due to a cancellation between dispersive and nonlinear scattering processes that add or remove a single photon from the quantum soliton mode. To see this, it is useful to use, in both formalisms, the system-reservoir expansion from Eq.~\eqref{eq:H_SR}.  Note that although $\hat{V}_1$ and $\hat{V}_2$ were defined in the ME formalism, they can also be written in the LSM formalism by writing $\hat{c}_z =\sum_{n\geq 1}u_n(z)\hat{a}_n$. As previously discussed, the uncoupled Hamiltonian $\hat{H}_{\rm S} + \hat{H}_{\rm R}$ recovers the classical soliton solution in the mean-field limit ($\bar{n}\rightarrow \infty$). Beyond this, as the reservoir begins in vacuum, the leading-order contribution to the destabilization of the soliton is given by $\hat{V}_1$. In the LSM formalism, $\hat{V}_1$ has a dispersive contribution of the form $\propto (\hat{a}_0\hat{a}_2^\dagger+\hat{a}_0^\dagger\hat{a}_2)$ because of the locality of GVD-induced couplings. For the nonlinear contribution, we can use the fact that $f_{\bar{n}}(z)=\sqrt{\bar{n}}u_0(z)$ satisfies the NLSE and evaluate it at $t=0$ to rewrite $u_0^3(z)$; this gives
\begin{equation}
C_{00}^{0m} = C_{00}^{m0}=C_{0m}^{00}=C_{m0}^{00} = \frac{D_{0m}}{2\bar{n}} +\frac{\bar{n}}{16}\delta_{0m}.
\end{equation}
As a result, $\hat{V}_1$ takes a simple form only involving $m=0$ and $m=2$ supermodes
\begin{align}
    \label{eq:h1}
    \hat{V}_1=D_{02}\left[1 - \frac{\hat{a}_0^{\dagger}\hat{a}_0}{\bar{n}}\right]\hat{a}_2^{\dagger}\hat{a}_0 + \text{H.c.},
\end{align}
where $D_{02} = \sqrt{5}\bar{n}^2/30$.
Notably, $\hat{V}_1$ vanishes upon substituting $\hat{a}_0\mapsto \sqrt{\bar{n}}$, suggesting that in the mean-field (classical) approximation, the soliton is unperturbed by $\hat{V}_1$. Fock states of the soliton with $\bar{n}+1$ photons are also eigenstates of $\hat{V}_1$ and stable to this order of coupling. As shown in Fig.~\ref{fig:supermodes}, this suppression of loss is to due a destructive interference between the linear coupling and the nonlinear coupling. 
In the ME formalism, as previously mentioned, a similar cancellation can be seen manifest in the form of $\mathcal{L}_1\hat{\rho}$ and its simplified $\hat{H}_{\rm fluc}$ form, which both arise from the coupling $\hat{V}_1$ and are stable in the same mean-field and Fock state sense.

The lowest-order scattering processes between the soliton and reservoir  reveal the mechanism behind the quantum mechanical stability of the soliton, and their contribution vanishes in the classical, mean-field limit. However, quantum fluctuations about the mean-field (or, equivalently, Fock state components with photon numbers deviating from $\bar{n}+1$) lead to deviations from this classical behavior. We find that the leading-order effects arise from: (i) fluctuations beyond the mean-field of $\hat{V}_1$, and (ii) the contribution of $\hat{V}_2$ which couples the soliton subspace to a scattering process involving two reservoir photons.

An estimate of the scaling of the effect of $\hat{V}_1$ can be found by assuming fluctuations of the soliton photon number consistent with an initial coherent state $\sim \sqrt{\bar{n}}$. The magnitude of a typical non-zero transition matrix element of $\hat{V}_1$ can thus be found by means of the scalings $\hat{a}^{\dagger}_2 \sim 1$, $\hat{a}_0 \sim \sqrt{\bar{n}}$, leading to $\hat{V}_1 \sim \bar{n}^2$, in contrast to a $\bar{n}^{\frac{5}{2}}$ scaling which would occur without the cancellation effect. In our simulations, we find $\hat{V}_1$ to primarily contribute to a quantum-induced phase shift of the soliton, with little effect on soliton evaporation (photon loss through photons leaving the fundamental soliton supermode).
 $\hat{V}_2$ similarly scales as $\sim\bar{n}^2$, which can be seen from inspection of its LSM representation, $\hat{V}_2 = -\sum_{nm}C_{nm}^{00}\hat{a}_0^2\hat{a}_n^{\dagger}\hat{a}_m^{\dagger} + \text{H.c.}$ Our simulations and ME theory indicate that $\hat{V}_2$ contributes an additional (small) nonlinear phase shift, and is also by far the dominant source of soliton evaporation. This loss scales for short times with $t^2/T_0^2$. This scaling and the origin of evaporation as coming from emission of entangled photon pairs are consistent with previous work~\cite{PhysRevA.98.043859, PhysRevA.107.053513}.

It is important to note that the couplings induced by $\hat{V}_1$ and $\hat{V}_2$ are \emph{dispersive} and diffusive, not resonant. A large energy gap is present between the soliton, which has approximate energy $-\bar{n}^3/8$, as seen from the classical solution, and the reservoir, which, for example in the LSM approach can be seen as set of modes indexed by $n$ with approximate energy $\bar{n}D_{nn}$ (as the two-photon component is negligible), which are positive and scale with $\bar{n}^3$. The couplings induced by $\hat{V}_1$ and $\hat{V}_2$ are thus highly off-resonant and---in terms of photon loss---contribute only to dispersive and \emph{non-Markovian} photon evaporation of the soliton~\cite{Gustin2025Feb}---henceforth, we refer to this as ``dispersive-dissipative'' photon loss. In contrast, the ME formalism captures only photon loss that arises from resonant couplings, giving rise to Markovian dissipation channels. This is thus a fundamental limitation of \emph{any} open quantum system formalism based on Markovian dissipation to capture quantum effects which lead to soliton evaporation.

Beyond the effects discussed above, there exist additional PFWM terms arising from $\hat{V}_{\rm others}$ (as well as potential non-Markovian effects not captured in the ME). These contributions are suppressed more rapidly with increasing $\bar{n}$ and we show them to be negligible in our simulations. 
\subsection{Numerical Results}\label{sec:numerics}
In this section, we perform numerical simulations to  verify and further investigate the qualitative considerations in Sec.~\ref{sec:instability}. 
For all simulations,  we assume the initial state $\ket{\psi_{\rm CS}}$, and transform to a frame rotating at $-\bar{n}^2/8$, which corresponds to the classical soliton phase evolution.

\subsubsection{Low-photon number regime}
We first investigate the low-photon number regime, where we perform full quantum simulations with a standard numerical package~\cite{Kraemer2018}. We use the LSM method with various finite truncations of number of supermodes, as well as the full ME given by the dissipator in Eq.~\eqref{eq:dissipators}, and these results are compared to a numerically-exact simulation with MPS~\cite{Yanagimoto2021Oct}, where we find convergence with a bond dimension of $50$ for all simulations. This regime corresponds to a scenario with high single-photon nonlinearities, where quantum fluctuations dominate and the soliton is strongly destabilized due to interactions beyond the mean-field classical stability.

    \begin{figure}[htb]
\includegraphics[width=1\columnwidth]{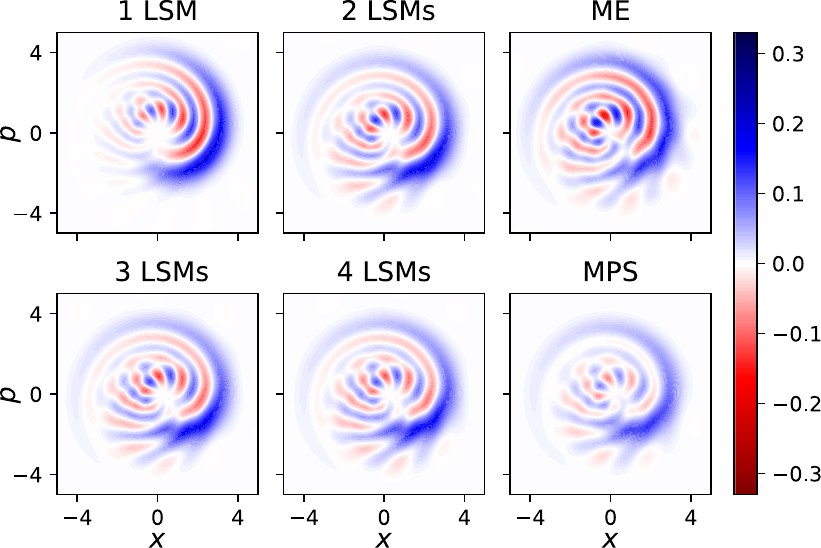}
    \caption{  Wigner function of the fundamental soliton supermode with $\bar{n}=5$ at time $t = 2T_0$. Simulation results
using the LSM and (single-mode) ME approaches are compared to the
full-quantum simulation using MPS (numerically exact). 
    }\label{fig:wigner}
\end{figure}

In Fig.~\ref{fig:wigner}, we show the Wigner function of the fundamental soliton supermode (mode index $n=0$) with initial photon number $\bar{n}=5$, where we observe large non-classical negativities. A truncation involving the three lowest LSMs suffices to capture the essential dynamical features, with the inclusion of higher supermodes making qualitatively minor contributions. 
The LSM method achieves a significant advantage over MPS-based simulation in terms of computational time, 
showing the efficacy of the LSM basis in extracting the essential soliton dynamics even with small $\bar{n}$. We also show the solution using the ME, which qualitatively reproduces the phase spreading of the full solution excellently, while not capturing the dispersive-dissipative photon loss. Being a single-mode theory, it is highly computationally efficient.

\begin{figure}[ht]
    \centering
    \includegraphics[width=1\columnwidth]{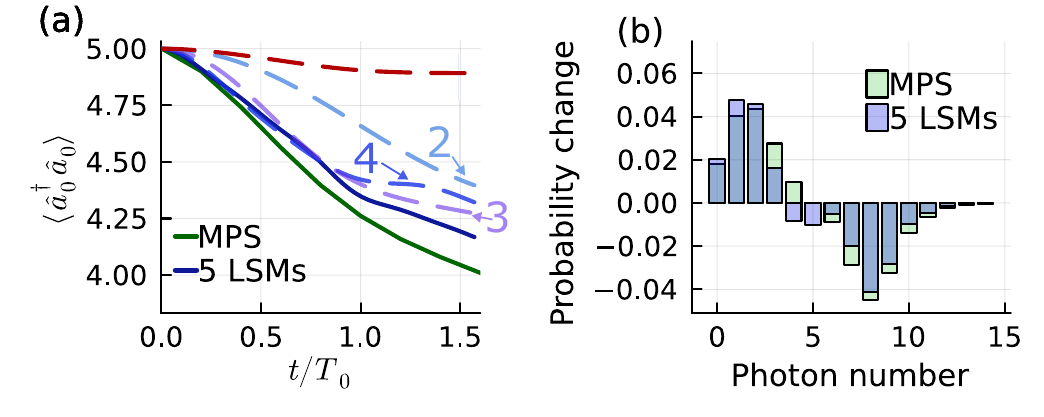}
    \caption{Simulations of a quantum $\chi^{(3)}$ soliton with $\bar{n}=5$. (a) Photon population in the soliton supermode as a function of time for LSM (with varying supermode number; see labels) and MPS  theory; the dashed red line corresponds to the 5 LSM model with $\hat{V}_2$ set to zero. (b) Change in the Fock state probability amplitude in the reduced $m=0$ density matrix at time $t = 1.4T_0$ compared with $t=0$ for 5 LSMs and MPS. 
    }\label{fig:low_n}
\end{figure}

To study the dispersive-dissipative soliton photon loss, we show in Fig.~\ref{fig:low_n}(a) the photon number decay for the LSM and MPS approaches. To show the strong effect of the $\hat{V}_2$ term of the dispersive-dissipative photon loss, we also plot the LSM theory calculation with the $\hat{V}_2$ removed from the Hamiltonian (note MPS can not remove $\hat{V}_2$ in the same way due to the requirement of spatially local couplings), revealing much stronger loss under the presence of the 2-reservoir-photon coupling from $\hat{V}_2$. While the LSM theory appears to converge slowly towards the MPS result, it captures the general trend well.  Figure~\ref{fig:low_n}(b) shows the photon number histograms of the change in relative Fock state probability $P_{n}(t) - P_n(0)$, where $P_n = \bra{n}\hat{\rho}^{\rm r}_{0}\ket{n}$, with $\hat{\rho}^{\rm r}_0 = \text{Tr}_{m\geq1}(\hat{\rho}_S)$, revealing photon loss away from the center quasi-stable point with $n \approx \bar{n}$ photons, suggestive of soliton formation.

\subsubsection{High-photon number regime}
Next, we move to a regime with larger $\bar{n}$, where the soliton becomes asymptotically closer to its stable classical solution, and we can employ a nonlinear Gaussian approximation to solve for the quantum mean-field and covariances~\cite{SI}.
For these large $\bar{n}$ simulations, we use the effective Hamiltonian form of the ME given by $\hat{H}_{\rm eff}$ [with $\hat{H}_{\rm fluc}$ defined in Eq.~\eqref{eq:Hfluc}]. This form not only facilitates the use of the nonlinear Gaussian state approximation, but also admits an exact analytical solution, as $\hat{H}_{\rm eff}$ is diagonal in the Fock state basis. We note there is no noticeable difference in our simulations with or without the nonlinear dispersive term proportional to $\Omega_{\bar{n}}$ arising from $\mathcal{L}_2$, indicating it is insignificant in this regime.

    \begin{figure*}[!htb]
    \centering
    \includegraphics[width=1.35\columnwidth]{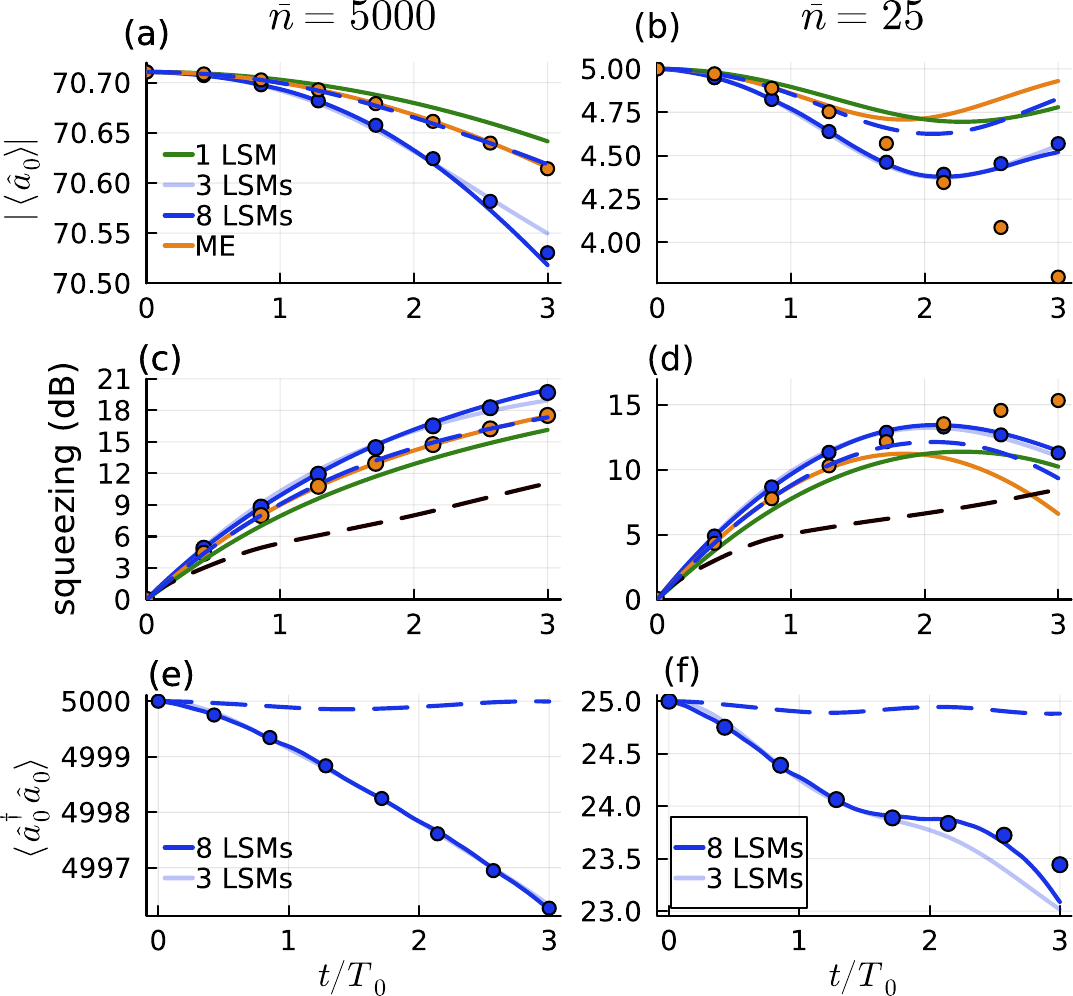}
    \caption{Simulations of solitons with the nonlinear Gaussian approximation. (a,b) soliton amplitude $|\langle \hat{a}_0 \rangle|$,   (c,d) maximum squeezing in the higher $\bar{n}$ regime, and (e,f) soliton photon number $\langle \hat{a}^{\dagger}_0 \hat{a}_0 \rangle$. The solid (dashed) blue lines correspond to a multimode LSM Gaussian model with (without) the $\hat{V}_2$ term and $N_{\rm LSM} = 8$, while the blue circles are calculated using the GSSF discretized continuum theory. We also show results for $N_{\rm LSM}=1$ and $N_{\rm LSM}=3$---note the 3 LSM curve often overlaps with the 8 LSM curve and thus is not always visible. The orange lines (circles) use the ME solution with (without) the Gaussian approximation. 
    In (c,d), the curves correspond to the squeezing supermode with the largest squeezing value; the dashed black line shows the squeezing value of the squeezing supermode with the second highest squeezing in the $N_{\rm LSM}=8$ model.
    }\label{fig:high_n}
\end{figure*}

In Fig.~\ref{fig:high_n}, we compare the LSM and ME approaches for large photon number $\bar{n}=5000$ and an intermediate regime $\bar{n} = 25$. In (a,b) the mean-field value over time of the fundamental soliton supermode $|\langle \hat{a}_0 \rangle|$  is shown (c.f. classical solution $|\langle \hat{a}_0\rangle| = \sqrt{\bar{n}}$) for the ME and LSM theories with the nonlinear Gaussian approximation.  Also shown is the exact ME solution,  as well as a Gaussian approach based on a spatial discretization of the field equations~\cite{SI}, where we have projected onto the fundamental supermode profile to express results in terms of expectation values of soliton $\hat{a}_0$, $\hat{a}_0^{\dagger}$ operators. 

By comparing the exact ME solution to the Gaussian solution, we see that deviations between the two models can occur on the order of the soliton period $T_0$, but are suppressed at higher photon numbers; at $\bar{n}=25$, very large deviations occur around $t \sim T_0$, whereas for $\bar{n}=5000$ only small deviations are seen even at $t \sim 3T_0$. In all cases, the LSM approach gives very similar results to the spatially discretized GSSF method, but uses a far smaller number of modes; here, the LSM method is very accurate even with $N_{\rm LSM}=3$, indicating that the seemingly complicated quantum behavior of the soliton can in fact be well-described in a concise few-mode basis.
 As with earlier, we also show results from the LSM model with $\hat{V}_2=0$. These reveal almost complete agreement with the ME model in the regime where the Gaussian approximation holds for the ME model, suggesting that cause for deviations with the full solution are almost entirely due to the neglect of the dispersive-dissipative contribution of the 2-reservoir-photon term. %As the Gaussian approximation is the only approximation made in the GSSF method, the validity of our results depend on the validity of this assumption. By comparison with the ME model Gaussian approximation, we see that it appears highly accurate for the timescales considered here.

In Fig.~\ref{fig:high_n}(e,f) we consider the photon loss from the fundamental soliton supermode using the Gaussian LSM and GSSF methods. For $\bar{n}=25$, the influence of potential Rabi oscillations with other LSMs can be seen, but this effect is suppressed in the $\bar{n}=5000$ case.  
In Fig.~\ref{fig:high_n}(c,d) we also consider the maximum squeezing of the quantum states during time evolution---for the multimode case, we decompose the system into its squeezing supermodes using the method from Ref.~\cite{ng2023quantumnoisedynamicsnonlinear} and plot the first and second of these maximum squeezing parameters, with good agreement seen between the ME and LSM models  only when $\hat{V}_2$ is neglected.

    \begin{figure}[htb]
    \centering
    \includegraphics[width=0.85\columnwidth]{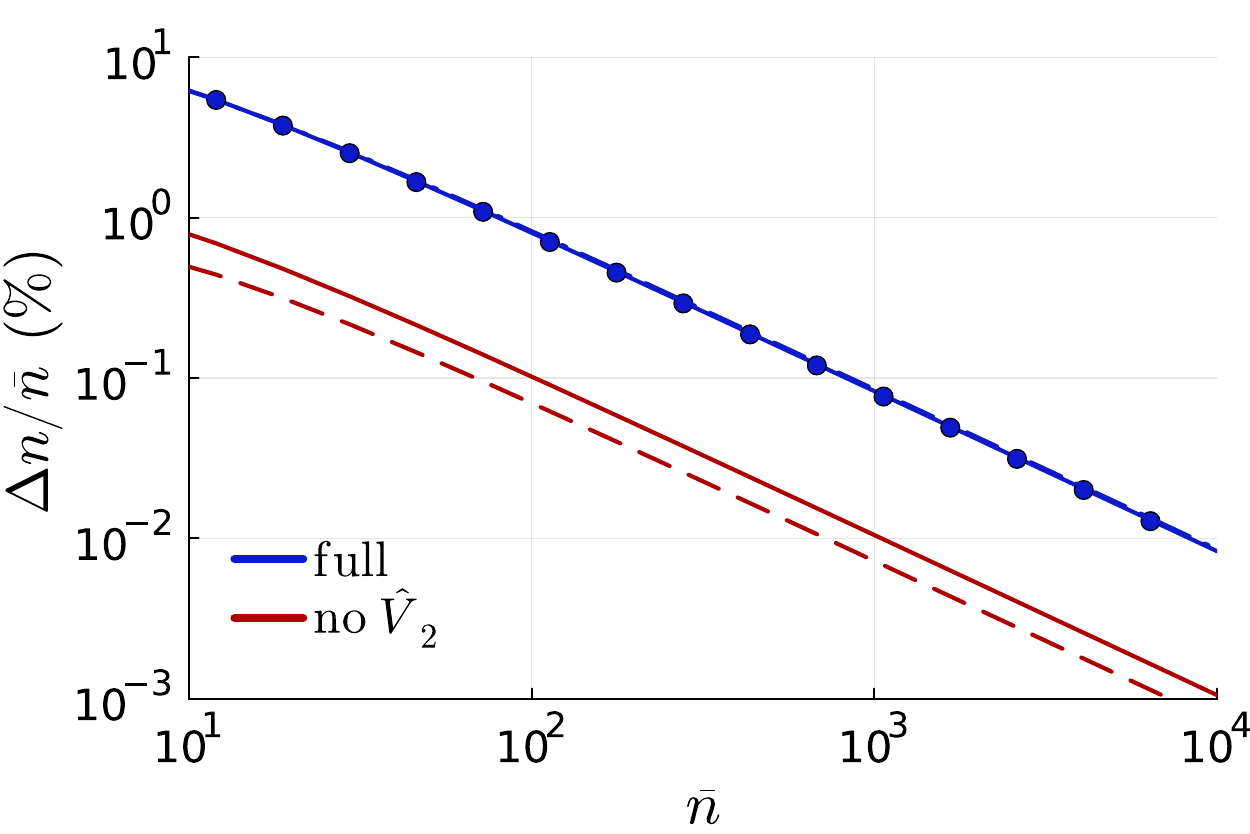}
    \caption{Fractional photon loss from the soliton supermode after one soliton period $t = T_0$ as a function of classical soliton photon number $\bar{n}$ with the nonlinear Gaussian approximation. The solid lines show the LSM model with $10$ supermodes, whereas the dashed  lines show $3$ supermodes. The blue lines are the full model (note solid and dashed are overtop of each other), and the red lines have $\hat{V}_2=0$. The blue dots correspond to the discretized continuum model. %
    }\label{fig:var_n}
\end{figure}

In Fig.~\ref{fig:var_n} we further investigate the role of $\hat{V}_2$ in leading to photon loss $\Delta n \equiv \langle \hat{a}_0^{\dagger}\hat{a}_0 \rangle - \bar{n}$ from the soliton. For all values of $\bar{n}$, the cause of photon loss is overwhelmingly from $\hat{V}_2$. Additionally, we again see that the full LSM approach converges very rapidly with even a small amount of supermodes ($N_{\rm LSM}=3$). The approximate slope of $-1$ in this log-log plot indicates that the number of photons loss $\Delta n$ after a soliton period is essentially independent of initial photon number under the nonlinear Gaussian approximation.

\section{Radiation into Dispersive Waves due to Higher-Order Dispersion}\label{sec:higher-order}

 The presence of a small higher-order dispersion can lead to radiative decay of the soliton through the production of dispersive waves~\cite{Akhmediev1995Mar}, which is a Markovian dissipation process capable of being described by the ME formalism~\cite{Gustin2025Feb}. This dispersion is captured by the $\hat{H}_{N}$ term from Eq.~\eqref{eq:H_fund_1}, which we neglected in our formalism until now for simplicity, but has been treated in the SI~\cite{SI}. We note that this effect is fundamentally classical in origin, in contrast to the uniquely quantum effects we have studied in the rest of the paper. 

This loss occurs due to a resonance condition arising in the tails of the soliton in reciprocal space between the reservoir and the main soliton mode. We consider the case $N=3$ here as an example. Using the full ME as derived in the SI~\cite{SI}, we find that dissipation occurs at values of $x = \pi k/\bar{n}$ which satisfy the phase-matching condition 
\begin{equation}\label{eq:disp}
    x^2 - \beta_3 x^3 - \tilde{\omega}_n =0, \end{equation}
    where $\beta_3 = -i2\bar{n} \eta_3/\pi \alpha_3$
    is the scaled (real-valued) TOD parameter. While it is possible to derive a ME that is valid for arbitrary $\bar{n}$ using the Fock transition operators, for simplicity we consider here only the large $\bar{n}$ regime where $\tilde{\omega}_n \approx -\pi^2/4$, and the phase-matching equation has one real solution at $x_0 =1/\beta_3 + \mathcal{O}(\beta_3)$. At these wavevectors, photons are emitted from the tails of the soliton and propagate rapidly away from the soliton due to their group-velocity mismatch, which leads to Markovian dissipation~\cite{Gustin2025Feb}. The full ME for this process can be described by adding to $\hat{H}_{\rm eff}$ (the effective Hamiltonian in the ME theory for large $\bar{n}$) the additional Hamiltonian $\hat{H}_{\rm fluc}^{(3)}$, which has the form
\begin{equation}
    \hat{H}_{\rm fluc}^{(3)} = \omega_{\rm L}\hat{a}_0^{\dagger}\hat{a}_0 + \omega_{\rm NL}\hat{a}_0^{\dagger 3}\hat{a}_0^3 + \omega_{\rm sol}\hat{a}_0^{\dagger}\left(1-\frac{\hat{a}_0^{\dagger} \hat{a}_0}{\bar{n}}\right)^2\hat{a}_0,
\end{equation}
as well as the Lindblads
\begin{subequations}
\begin{align}
    \hat{L}_{\rm L} &= \sqrt{\gamma_{\rm L}}\hat{a}_{0}, \\
        \hat{L}_{\rm NL} &= \sqrt{\gamma_{\rm NL}}\hat{a}_0^{\dagger} \hat{a}_0^2, \\
        \hat{L}_{\rm sol} &= \sqrt{\gamma_{\rm sol}}(1-\frac{\hat{a}_0^{\dagger}\hat{a}_0}{\bar{n}})\hat{a}_0,
    \end{align}
\end{subequations}
where all $\omega$ and $\gamma$ coefficients are functions of $\beta_3$ given in~\cite{SI}. To leading order in their contribution to the ME for small $\beta_3$ and large $\bar{n}$, they take the approximate effective forms $\omega_{\rm sol} \approx \gamma_{\rm sol} \approx \gamma_{\rm L} \approx \omega_{\rm L} \approx 0$, and
\begin{subequations}\begin{align}\label{eq:nl_gamma}
        \gamma_{\rm NL} &\approx \frac{2}{\pi |\beta_3|^3}e^{-\frac{2}{|\beta_3|}}, \\
        \omega_{\rm NL} &\approx \left(\frac{\pi^4}{384} - \frac{17\pi^2}{480}\right)\beta_3^2
    \end{align}
    \end{subequations}
which is also consistent with the classically predicted decay rate~\cite{Akhmediev1995Mar}, at least up to a possible numerical prefactor. That the most dominant effects are a nonlinear dissipation channel and corresponding cascaded (effective) $\chi^{(4)}$ SPM can be understood by noting that the nonlinear interaction in the soliton supermode is required to lose photons through a 1-reservoir photon coupling channel while still conserving momentum, and the corresponding phase shift arises generally as a result of Kramers-Kronig like causality~\cite{Onodera2022Mar}.

\begin{figure}[htb]
    \centering
    \includegraphics[width=0.99\columnwidth]{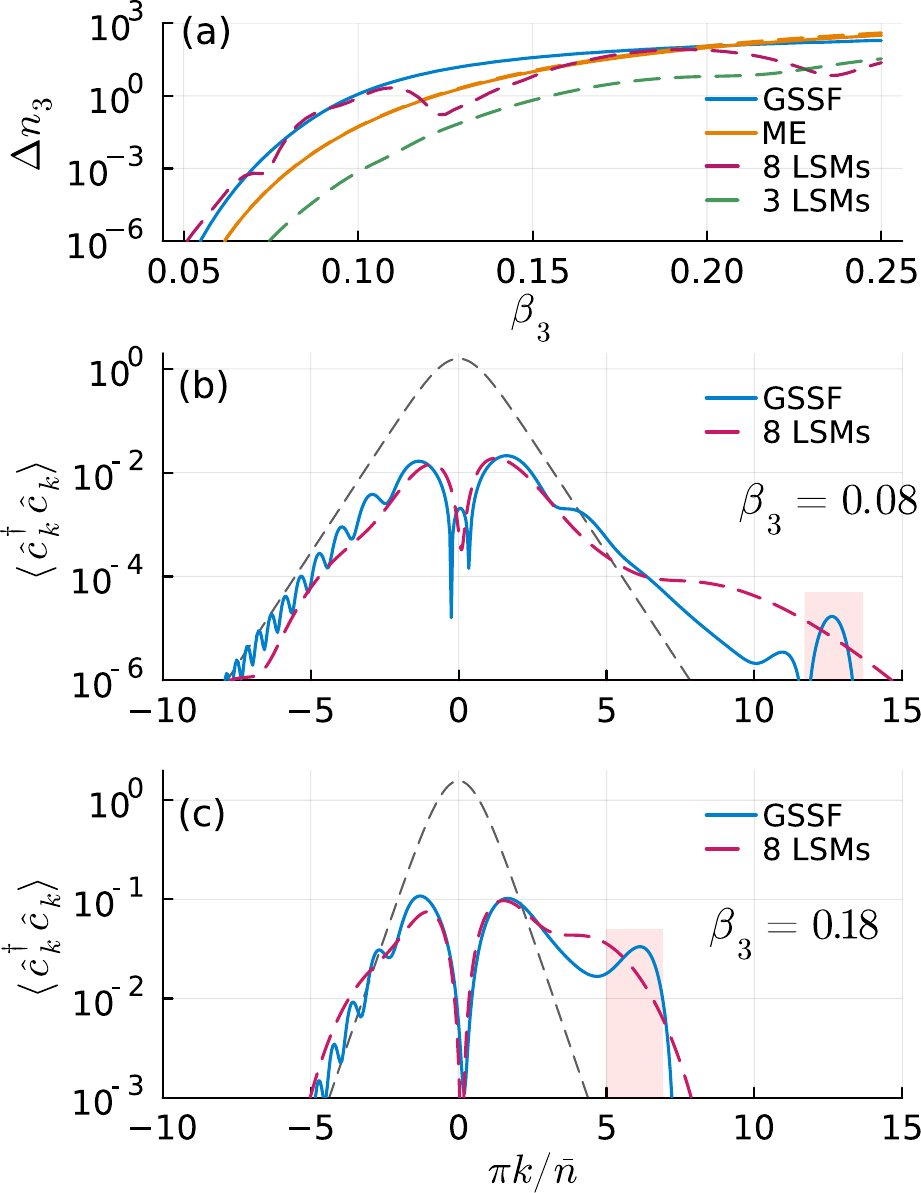}
    \caption{(a) dispersive wave photon loss $\Delta n_3$ from a $\bar{n}=5000$ soliton as a function of TOD parameter $\beta_3$ for GSSF, ME, and LSM models, with window width $\pi\delta k / \bar{n} =1$ and at time $t= 2T_0$. For the ME model we approximate $\Delta n_3 \approx \bar{n} - \langle \hat{a}^{\dagger}_0 \hat{a}_0 \rangle$. The approximate form given in the text with only the rates of Eq.~\eqref{eq:nl_gamma} is plotted as solid orange, and the full~\cite{SI} ME is given by dashed orange lines.  (b,c)  reservoir photon number distribution for the simulation in (a) using two specific values of $\beta_3$, with window around $k_0 \pm \delta k$ shaded in light red. An outline of the $\tilde{u}_0(k)$ supermode is shown as a dashed black line for reference. 
    }\label{fig:dispersion}
\end{figure}

In Fig.~\ref{fig:dispersion}(a), we plot the dispersive wave soliton photon loss as a function of TOD parameter $\beta_3$ for the discretized GSSF, ME (approximate and full), and LSM models. We use the nonlinear Gaussian approximation for all. We define the dispersive wave loss as $
    \Delta n_{3} \equiv \int_{k_0 -\delta k}^{k_0 + \delta k} dk \langle \hat{c}_{k}^{\dagger} \hat{c}_k \rangle$,  where
$\hat{c}_k = \frac{1}{\sqrt{2\pi}}\intinf dz e^{ikz} \hat{c}_z$, and $\hat{c}_z = \hat{\phi}_z - u_0(z) \hat{a}_0$. In the LSM approach, $\hat{c}_z = \sum_{n\geq 1}u_n(z)\hat{a}_n$. The reservoir photon number is integrated over an emission window centered around $k_0 = \bar{n} x_0/\pi$, where $x_0$ is the real solution to Eq.~\eqref{eq:disp}, with window width $2\delta k$. We note that strictly speaking, for comparison with loss from the soliton supermode, the definition of $\Delta n_3$ would also include cross-terms of the soliton supermode and reservoir, but these should be negligible for dissipation localized far from the soliton profile.
    
    While the ME theory predicts the correct dissipation trend, it dramatically underestimates the amount of dissipation, except at larger $\beta_3 \gtrsim 0.2$, where the substantial fractional photon loss makes the soliton basis questionable. The LSM theory also performs poorly, with oscillatory behavior seen in the $N_{\rm LSM}=8$ case which only partially captures on average the correct trend, and $N_{\rm LSM}=3$ failing completely. We also note the ME using the approximate rates from Eq.~\eqref{eq:nl_gamma} overlaps excellently with the full ME solution.

To understand these results, we plot in Fig.~\ref{fig:dispersion}(b,c) the reservoir photon number distribution $\langle \hat{c}_k^{\dagger} \hat{c}_k \rangle$. We observe a reservoir emission peak around the phase-matching condition $\pi k/\bar{n} \approx 1/\beta_3$. It can be clearly seen here why the LSM theory fails to capture the correct trend and instead oscillates with very large amplitude around it; the LSM theory is constrained to $N_{\rm LSM}$ degrees of freedom in its shape in reciprocal space, and thus can not accurately resolve the sharp resonance that occurs at the phase-matching condition, except with a very large number of supermodes. 
Additionally, we observe significant \emph{broadening} of the photon profile due to broadband reservoir photon emission on the side of the soliton which contains the resonance, which we attribute to dispersive-dissipative interactions.

This latter effect gives a potential explanation for the underestimate of the ME (and thus classical) dissipation prediction. In the ME theory, the effective density of states (DOS) for the dissipative photon emission process is proportional to the intensity of the Fourier-transformed soliton supermode $\tilde{u}_0^2(k)$, where $\tilde{u}_0(k) = \frac{1}{\sqrt{2\pi}}\intinf dz e^{ikz}u_0(z) = \sqrt{\frac{\pi}{2\bar{n}}}\text{sech}(\pi k/\bar{n})$ [which gives rise to the exponential factor in Eq.~\eqref{eq:nl_gamma}]. While the ME theory retains only the $n=0$ supermode, if higher order LSMs were also included into the theory, these would have dissipation rates using instead $\tilde{u}_n^2(k)$ as their effective DOS. Since these are broader, for small $\beta_3$, these modes will exhibit much larger dissipation; for example, $\tilde{u}_1^2(k_0)/\tilde{u}_0^2(k_0) \approx 12/(\pi^2\beta_3^2) \gg1$, and $\tilde{u}_2^2(k_0)/\tilde{u}_0^2(k_0) \approx 45/(\pi^4 \beta_3^4) \ggg 1$. We thus propose the following mechanism for the acceleration of photon loss beyond the single-supermode ME and classical prediction observed for small $\beta_3$: dispersive-dissipative couplings induce transitions between the soliton and higher-order LSMs, which rapidly exhibit Markovian photon dissipation due to the phase-matching condition. Additionally, the loss in photons from the soliton supermode destabilizes the soliton stability (c.f. the form of $\hat{V}_1$), leading to further enhancement of higher-order LSM population and thus dissipation.

\section{Conclusions and Outlook}\label{sec:conc_outlook}

In this section we briefly discuss prospects for experimental observation of the effects we predict in this work. We then conclude, and discuss possible extensions of our work and future study.

\subsection{Experimental Prospects}

While our results have been stated in nondimensionalized units, a connection to observable parameters can be found by bringing back in the dimensionful forms~\cite{SI}. In particular, we can find the relationship
\begin{equation}\label{eq:T0rel}
    \tilde{T}_{0} = 0.51\frac{T_{\rm FWHM}^2}{v_g |k''|},
\end{equation} 
where $\tilde{T}_0$ is the (temporal) soliton period in physical units. 
Equation~\eqref{eq:T0rel} allows one to relate our results, which have been expressed temporally in terms of the soliton period, to experimentally observable quantities for a given soliton. The SI also contains an expression for the soliton period in physical units in terms of photon number~\cite{SI}.

In terms of dispersive wave radiation due to third-order dispersion, the dimensionless parameter $\beta_3$ can be related to physical quantities in dimensionful units by means of~\cite{SI}
\begin{equation}\label{eq:beta3_rel}
\beta_3 = -\frac{1.1 v_g |k''|}{T_{\rm FWHM}}\left(1 - \frac{k'''}{3v_g k''^2}\right),
\end{equation}
where $T_{\rm FWHM}$ is the full width at half maximum of the initial spatial soliton intensity profile, and $v_g$ and $|k''|$ are the group velocity and GVD of the medium, respectively. $k'''$ is the TOD parameter of the medium.
We can obtain estimates of when the TOD will roughly become significant by letting $\beta_3 \approx 0.1$ [see Fig.~\ref{fig:dispersion}(a)]. For solitons in optical fibers, we use characteristic numbers~\cite{PhysRevA.103.013521} $v_g = 2 \times 10^8$ m/s, $k'' = -20 \ \text{ps}^2/\text{km}$ and $k''' = 0.1 \ \text{ps}^3/\text{km}$, finding $\beta_3 \approx 1.8\left(T_{\rm FWHM}/1 \ \text{fs}\right)^{-1}$,
and so the TOD becomes significant for pulses less than $\sim 18 \text{ fs}$. For silicon nitride---a material which boasts high $\chi^{(3)}$ nonlinearities---using example waveguide numbers~\cite{tan_group_2010} $v_g \approx 1.4 \times 10^8$ m/s, $k'' \approx -500 \ \text{ps}^2/\text{km}$, and $k''' \approx 2 \ \text{ps}^3/\text{km}$, which gives $\beta_3 \approx 1.3\left(T_{\rm FWHM}/1 \ \text{fs}\right)^{-1}$, and TOD becomes significant below around $13$ fs.

An interesting prospect would be to observe the non-Gaussian behavior that our formalism is capable of studying in an experiment. This is typically very challenging, as it requires the soliton period in the few photon regime to be comparable or shorter than the characteristic photon loss rate. While experimentally, third-order optical nonlinearities are typically weak enough to render such non-Gaussianities negligible, we stress that our formalism could also be applied to a variety of other solitonic systems, including Bose-Einstein condensates~\cite{Khaykovich2002} and $\chi^{(2)}$ solitons (known as ``simultons'')~\cite{Yanagimoto2021Oct,BURYAK200263}. In particular, recent advances in $\chi^{(2)}$ platforms have suggested that the regime of substantial coupling in the few-photon regime may become experimentally accessible in the near future~\cite{Hickstein2019Jul,Jankowski2021Sep,Mishra2022Aug,Singh2020,Nehra2022few,Solntsev2022Jun,Yanagimoto2024Jul}. 
 $\chi^{(2)}$ simultons accumulate phase according to $\exp(i\phi_0 t/3)$ with $\phi_0=(3\bar{n}^2/32)^{1/3}$ for the soliton photon number of $\bar{n}$ in normalized units~\cite{Yanagimoto2021Oct}. Here, $t=1$ corresponds to a propagation distance of $L_{\chi^{(2)}}=(k''/(\hbar \omega_\text{SH}\eta))^{1/3}$, where $k''$ is the signal GVD, and $\eta$ is the normalized SHG conversion efficiency of the waveguide. Using the same definition as the Kerr case, the simulton ``period'' should be $\frac{2\pi}{3(3n^2/32)^{1/3}}$. For $n=100$ photons, this number is about $0.2$, meaning we may observe non-Gaussian physics around $t=0.2$.  Using the state-of-the-art numbers  (see Ref.~\cite{Yanagimoto2022Apr}), we have $L_{\chi^{(2)}}=1.4\,\text{cm}$ on lithium niobate. 
%I think even conservative numbers will give us $L_{\chi^{(2)}}=1\,\text{m}$. 
 On the other hand, a characteristic attenuation distance of the state-of-the-art nanophotonics is about 1 m~\cite{Sham-Ansari2022}. This suggests that the kind of non-Gaussian physics we discuss in this work will be relevant in $\chi^{(2)}$ systems. 
We expect our formalisms to serve as indispensable tools for efficiently studying the uniquely quantum effects associated with soliton-like physics in such regimes.

\subsection{Conclusions}
We have presented two new powerful methods to modeling quantum soliton propagation that are capable of going beyond the linearization regime, the LSM and ME approaches. Both methods begin with projecting the fundamental (classical) soliton spatial profile onto the underlying quantum field; the LSM approach then treats the residual component of the field as a discrete expansion over LSMs generated through an orthonormalization procedure, whereas the ME approach leaves the residual component as a continuum field, imparted with non-local correlations. We have shown that these approaches offer both physical insight into how the stability of the soliton manifests in the quantum pictures, and how quantum fluctuations lead to a hierarchy of destabilizing higher-order effects, but from different perspectives. We have also shown how the ME theory can recover (and go beyond) classically known expressions for the dispersive wave radiation rates, and identified how these significantly underestimate the actual amount of radiation due to dispersive-dissipative broadening of the soliton, which increases the effective DOS of the radiative process. These methods should prove to be a powerful tool of analysis of soliton physics, due to their ability to constrain complex and non-Gaussian quantum dynamics effectively within a small subspace of one or more dominant quantum degrees of freedom.

A particularly powerful application of our method is to treat the non-Gaussian mesoscopic regime, where the soliton can consist of of dozens-to-hundreds of photons~\cite{Yanagimoto2024Jul}. We have shown that our ME formalism can effectively model the quantum-induced phase fluctuations of the soliton in this regime, where most other known approaches are either based on linearization (which neglects non-Gaussianity), exact solutions (which can be difficult to work with and not generalizable to non-integrable systems), or are intractable with many photons.

Our work provides many opportunities for generalization and further research. We have seen throughout that a 3-mode LSM model can effectively describe much of the quantum soliton physics  (and the structure of the coupling matrices $D_{nm}$ and $C_{nm}^{n'm'}$ indicate a two mode model with $n=0$ and $n=2$ LSMs would be almost equally effective; c.f. Fig.~\ref{fig:coupling}). Combining the ME and LSM approaches by expanding the system as a dominant two (or more) supermode system basis and a residual reservoir and then deriving a ME would likely provide a powerful model able to capture both nonlinear phase shifts and dissipative photon loss across a wide parameter regime.
Such an approach holds promise especially in the non-Gaussian mesoscopic regime, where full LSM simulations with more than a very small number of supermodes can be highly challenging numerically. It would also be particularly interesting to study $\chi^{(2)}$ simultons using our approach. Based on state-of-the-art experimental numbers, this platform has a potential of approaching single-photon nonlinearities in a foreseeable future, where non-Gaussian quantum physics naturally emerge.

Finally, the ME approach based on partitioning of the quantum field that we have developed in this work is highly general, and we expect it to be a powerful tool for studying nonlinear quantum dynamics across a variety of different platforms where dynamics remain close-to-localized in a well-defined system basis.

\acknowledgments
 We thank NTT Research for their
financial and technical support. We thank Nuno Castanheira for useful conversations.

\section*{Funding}
This work was supported by NSF Awards CCF-2423832 and CCF-1918549.
\section*{Disclosures}
The authors declare no conflicts of interest.

    \bibliography{references}

\end{document}